\documentclass[12pt]{article}
\usepackage[mathcal]{euscript}
\usepackage{graphicx}
\usepackage{amssymb}
\usepackage{epstopdf}
\usepackage{latexsym}
\usepackage{indentfirst}
\def\Tr{{\rm Tr}}

\def\CD{{\cal D}}
\def\K3{\mathrm K3}

\def\double #1{#1{\hbox{\kern-2pt $#1$}}}

\def\half {{1\over 2}}

\def\Ab{ {\bf A}}
\def\Abd{ {\bf A}^\dagger}

\def\bb{ {\bf b}}
\def\cb{ {\bf c}}

\def\bbd{ {\bf b}^\dagger}
\def\cbd{ {\bf c}^\dagger}

\def\pp{{\mathchoice
          {%displaystyle
              \kern 1pt%
              \raise 1pt
              \vbox{\hrule width5pt height0.4pt depth0pt
                    \kern -2pt
                    \hbox{\kern 2.3pt
                          \vrule width0.4pt height6pt depth0pt
                          }
                    \kern -2pt
                    \hrule width5pt height0.4pt depth0pt}%
                    \kern 1pt
           }
            {%textstyle
              \kern 1pt%
              \raise 1pt
              \vbox{\hrule width4.3pt height0.4pt depth0pt
                    \kern -1.8pt
                    \hbox{\kern 1.95pt
                          \vrule width0.4pt height5.4pt depth0pt
                          }
                    \kern -1.8pt
                    \hrule width4.3pt height0.4pt depth0pt}%
                    \kern 1pt
            }
            {%scriptstyle
              \kern 0.5pt%
              \raise 1pt
              \vbox{\hrule width4.0pt height0.3pt depth0pt
                    \kern -1.9pt  %[e]=0.15pt
                    \hbox{\kern 1.85pt
                          \vrule width0.3pt height5.7pt depth0pt
                          }
                    \kern -1.9pt
                    \hrule width4.0pt height0.3pt depth0pt}%
                    \kern 0.5pt
            }
            {%scriptscriptstyle
              \kern 0.5pt%
              \raise 1pt
              \vbox{\hrule width3.6pt height0.3pt depth0pt
                    \kern -1.5pt
                    \hbox{\kern 1.65pt
                          \vrule width0.3pt height4.5pt depth0pt
                          }
                    \kern -1.5pt
                    \hrule width3.6pt height0.3pt depth0pt}%
                    \kern 0.5pt%}
            }
        }}

\def\mm{{\mathchoice
                       {%displaystyle
                             \kern 1pt
               \raise 1pt    \vbox{\hrule width5pt height0.4pt depth0pt
                                  \kern 2pt
                                  \hrule width5pt height0.4pt depth0pt}
                             \kern 1pt}
                       {%textstyle
                            \kern 1pt
               \raise 1pt \vbox{\hrule width4.3pt height0.4pt depth0pt
                                  \kern 1.8pt
                                  \hrule width4.3pt height0.4pt depth0pt}
                             \kern 1pt}
                       {%scriptstyle
                            \kern 0.5pt
               \raise 1pt
                            \vbox{\hrule width4.0pt height0.3pt depth0pt
                                  \kern 1.9pt
                                  \hrule width4.0pt height0.3pt depth0pt}
                            \kern 1pt}
                       {%scriptscriptstyle
                           \kern 0.5pt
             \raise 1pt  \vbox{\hrule width3.6pt height0.3pt depth0pt
                                  \kern 1.5pt
                                  \hrule width3.6pt height0.3pt depth0pt}
                           \kern 0.5pt}
                       }}

\def\ad{{\kern0.5pt
                   \alpha \kern-5.05pt
\raise5.8pt\hbox{$\textstyle.$}\kern 0.5pt}}

\def\bd{{\kern0.5pt
                   \beta \kern-5.05pt \raise5.8pt\hbox{$\textstyle.$}\kern 0.5pt}}

\def\qd{{\kern0.5pt
                   q \kern-5.05pt \raise5.8pt\hbox{$\textstyle.$}\kern 0.5pt}}
\def\Dot#1{{\kern0.5pt
     {#1} \kern-5.05pt \raise5.8pt\hbox{$\textstyle.$}\kern 0.5pt}}

\begin{document}

\setcounter{page}0
\thispagestyle{empty}
\begin{flushright}
\makebox[0pt][b]{YITP-SB-08-17}
\end{flushright}
\vspace{25pt}

\begin{center}
{\LARGE Integrability of the Gauged Linear Sigma Model for $AdS_5\times S^5$ }\\

\vspace{30pt}
{
William D. Linch,
III${}^{\clubsuit}$\footnote{email: wdlinch3@math.sunysb.edu} and Brenno Carlini
Vallilo${}^{\spadesuit}$\footnote{email: vallilo@unab.cl}
}
\vspace{10pt}

${}^{\clubsuit}${\em C.N. Yang Institute for Theoretical Physics\\ 
and\\
Department of Mathematics,\\ SUNY, Stony Brook, NY 11794-3840, USA}\\
~\\
~\\
${}^{ \spadesuit}${\em Departamento de Ciencias F\'{\i}sicas, Universidad Andres Bello \\ Republica 220, Santiago, Chile}
\vspace{70pt}

 {\bf Abstract} 
\end{center}
Recently, a gauged linear sigma model was proposed by Berkovits and Vafa
which can be used to describe the $AdS_5\times S^5$ superstring at
finite and zero radius. In this paper we show that the model is
classically integrable by constructing its first non-local conserved charge
and a superspace Lax ``quartet''. Quantum conservation of the non-local
charge follows easily from superspace rules.
%Using the superspace form of the non-local charge, we give a simple proof that the model is integrable at the quantum level.

\newpage

%\tableofcontents

%%%%%%%%%%%%%%%%%%%%%%%%%%%%%%%%%%%%%%%%%%%%%%%%%%
\section{Introduction}

Over the past ten years there has been much activity in the $AdS$/CFT correspondence. This powerful 
conjecture \cite{adscft} relates two different theories in different regimes. 
It is very difficult to prove the correspondence fully, since this would
involve a complete solution of the theories on both sides. Nevertheless, we would like to
see how the fundamental degrees of freedom on one side of the conjecture
appear on the other side.

One particular limit which could be interesting to analyze is the limit in which
the super Yang-Mills theory is free. Although we have a trivial theory on one side
of the conjecture, the dynamics of the string theory side is governed by a
highly interacting worldsheet. This limit is beyond the reach of perturbation
theory using the Metsaev-Tseytlin $AdS_5\times S^5$
Green-Schwarz sigma model \cite{metsaev} (or its pure spinor \cite{pure}\ version \cite{chan,oneloop}). Although both
versions appear to be integrable two-dimensional field theories
\cite{bena,flat,berk1}, no one has been
able to use integrability to perform a non-trivial calculation which could
shed light on the strongly-coupled regime. A possible approach to this problem was recently 
proposed by Berkovits and Vafa \cite{berkvafa}. Using a modified version
of the pure spinor action in $AdS_5\times S^5$, they were able to
define a gauged linear sigma model which is related to the usual
superspace variables by a twistor-like field redefinition. The model
so obtained has $N=(2,2)$ worldsheet supersymmetry, global $U(2,2|4)$,
and local $U(4)$ symmetry. The fact that the global symmetry group is a
supergroup has important implications for the quantum theory. After
integrating out the gauge degrees of freedom, one recovers the non-linear sigma model 
action previously obtained in \cite{berk}.

Although the original motivation in \cite{berkvafa} was to construct an
action in which the zero-radius limit is reachable, the non-linear sigma 
model action is supposed to be equivalent to
the pure spinor version for all radii. Furthermore, since it has
$N=(2,2)$ worldsheet supersymmetry and space-time supersymmetry it is
possible that many quantum calculations are greatly simplified. In this
work we show that this is indeed the case. 

A subtle
point is the definition of the physical spectrum. Although Berkovits
and Vafa refer to their model as an ``A-model'', the physical spectrum, which
is supposed to be equivalent to the pure spinor version, is not the usual
cohomology of an A-model since, the BRST charge of the pure spinor
description is not mapped to the BRST charge of the A-model. Only the low-lying 
excitations, which were used in \cite{berkvafa}, should agree using
the two different BRST charges.\footnote{The topological sector of the
sigma model was recently used in \cite{bonelli} to compute amplitudes in the open string
sector of $\half$-BPS operators.}  

In this paper we study the classical and quantum integrability of this gauged
linear sigma model. The worldsheet supersymmetry plays an important role
in constraining the form of possible quantum corrections in the effective
action and correlators, and space-time
supersymmetry helps to prove that many of these corrections vanish. The
end result is that the first non-local charge is a well-defined operator in 
the quantum theory and does not need renormalization. This provides
further evidence that the gauged linear sigma model picture is a
consistent description of the pure spinor superstring in $AdS_5\times
S^5$.

Integrability techniques are well developed on the YM side of the
conjecture where the full S-matrix \cite{matthiassmatrix,beisertsmatrix} and Bethe equations, 
which determine
the anomalous dimensions of gauge theory operators in the long operator
limit, was already derived \cite{minahan,staudacher1,staudacher2}. Also, a complete
anomalous dimension function of some particular gauge theory operator
which, was shown to agree with both perturbative YM \cite{bern} and string
theory \cite{roiban} sides, 
was constructed in \cite{staudacher2}. We hope that the high number of
space-time and worldsheet symmetries of this gauged linear sigma model
will facilitate the implementation of such a program on the
string theory side. It would be very interesting to see how the methods of 
\cite{kazakov1,beisert2,beisert3,gromov1,gromov2, mikhailov1,mikhailov2} can be applied to the present case.

This paper is organized as follows. In section \ref{Definition}, we introduce the gauged
linear sigma model proposed by Berkovits and Vafa. In section \ref{symmetries}, we discuss
its classical symmetries and find the corresponding non-local conserved
charges. Section \ref{classint} is devoted to the discussion of classical
integrability. In section \ref{quantumint}, we address the question of
quantum integrability of the sigma model. We conclude and discuss open
problems in section \ref{Conclusion}. In the appendix we put definitions
and derivations which were skipped in the main text.

%%%%%%%%%%%%%%%%%%%%%%%%%%%%%%%%%%%%%%%%%%%%%%%%%
\section{Definition of the GSLM}
\label{Definition}
The gauged linear sigma model defined by  Berkovits and Vafa \cite{berkvafa} is
related to the pure spinor $AdS_5\times S^5$ sigma model after a BRST-trivial 
term is added to the action. This BRST-trivial term enhances the
target space symmetries and makes it possible to describe the model in terms of an 
$N=(2,2)$ supersymmetric worldsheet action principle.

The resulting model resembles the old Grassmannian sigma models on
$\frac{U(n+m)}{U(n)\times U(m)}$ (see, {\it e.g.}~\cite{grassmann}), but we replace the numerator with the supergroup $U(2,2|4)$
(and also replace one of the $U(4)$s with $U(2,2)$).\footnote{For a review of supergroups, see appendix \ref{Sec:Supergroups}.} The second fundamental difference is that, by construction, the worldsheet
fields are fermionic and will have a kinetic term with two derivatives.
The choice of denominator makes the model a symmetric space, in
contrast with the $AdS_5\times S^5$ sigma model which also has a Wess-Zumino term. 

We begin by establishing some $N=(2,2)$ superspace notation. Bosonic worldsheet coordinates will be denoted by $(\sigma^\pp,\sigma^\mm)$ and the 
fermionic coordinates will be denoted by
$(\kappa^+,\kappa^-,\bar\kappa^+,\bar\kappa^-)$. The covariant
superderivatives are taken to be
\begin{eqnarray}
D_+ =\frac{\partial}{\partial\kappa^+} -i \bar\kappa^+\frac{\partial}{\partial \sigma^\pp}&,& 
		\bar D_+ =\frac{\partial}{\partial\bar\kappa^+} -i \kappa^+\frac{\partial}{\partial \sigma^\pp},  \cr
			&&\cr
D_- =\frac{\partial}{\partial\kappa^-} -i \bar\kappa^-\frac{\partial}{\partial\sigma^\mm}&,& 
	\bar D_-=\frac{\partial}{\partial\bar\kappa^-} -i \kappa^-\frac{\partial}{\partial\sigma^\mm}.
\end{eqnarray}
They commute with the supercharges
\begin{eqnarray}
\label{susy}
Q_+ =\frac{\partial}{\partial\kappa^+} +i \bar\kappa^+\frac{\partial}{\partial \sigma^\pp}&,& 
		\bar Q_+ =\frac{\partial}{\partial\bar\kappa^+} +i \kappa^+\frac{\partial}{\partial \sigma^\pp},  \cr
		&&\cr
Q_- =\frac{\partial}{\partial\kappa^-} +i \bar\kappa^-\frac{\partial}{\partial\sigma^\mm}&,& 
	\bar Q_-=\frac{\partial}{\partial\bar\kappa^-} +i \kappa^-\frac{\partial}{\partial\sigma^\mm},
\end{eqnarray}
and satisfy the anticommutation relations
\begin{eqnarray}
\{ D_+,\bar D_+\}= -2i\partial_\pp &,& \{ D_-,\bar D_-\}=-2i\partial_\mm,
\end{eqnarray}
where $\partial_\pp={\partial}/{\partial \sigma^\pp}$ and
$\partial_\mm={\partial}/{\partial\sigma^\mm}$. Any other graded commutator
vanishes. Integration over the full superspace is defined as
\begin{eqnarray}
\int d^4\kappa \, = D_+D_-\bar D_+\bar D_- \, \big|_{\kappa^+=\kappa^-=\bar\kappa^+=\bar\kappa^-=0}.
\end{eqnarray} 

Analogously to the bosonic Grassmannian \cite{grassmann} sigma models, we introduce
the basic fields
$\Phi^\Sigma_R(\sigma, \kappa)$. Here $\Sigma$ is a global $U(2,2|4)$ 
index which splits into $A=1,...,4$ and $J=1,...,4$, where $A$ is a bosonic global
$U(2,2)$ index, and $J$ is a fermionic global $U(4)$ index. $R$ is a fermionic local 
$U(4)$ index which will be gauged by introducing a gauge prepotential
$V^R_S(\sigma,\kappa)$. Note 
that since $U(2,2|4)$ is a supergroup, and $\Phi^\Sigma_R$ is in its fundamental
representation, $\Phi^A_R$ is a fermionic superfield and $\Phi^J_R$ is
a bosonic superfield.

The superfields come in chiral/anti-chiral pairs
\begin{eqnarray} \bar D_+ \Phi^\Sigma_R=\bar D_-\Phi^\Sigma_R=0,\quad  D_+
\bar\Phi^R_\Sigma=
D_-\bar\Phi^R_\Sigma=0,\end{eqnarray}
which have the following expansion in terms of component
fields:\footnote{This differs from the expansion in \cite{berkvafa} which
has the wrong component fields in the antichiral field.}
\begin{eqnarray}\label{fundfields}
\Phi^\Sigma_R= \phi^\Sigma_R + \kappa^+ X^\Sigma_R +\kappa^- \bar Y^\Sigma_R +
\kappa^+\kappa^- F^\Sigma_R + \cdots ,\nonumber \\
\bar \Phi_\Sigma^R= \bar\phi_\Sigma^R + \bar\kappa^+ Y_\Sigma^R
+\bar\kappa^- \bar X_\Sigma^R +
\bar\kappa^+\bar\kappa^- F_\Sigma^R + \cdots ,
\end{eqnarray}
where $X^\Sigma_R$ will (after fixing an appropriate gauge) be a twistor-like combination of the $AdS_5\times S^5$
coordinates and pure spinor ghosts, $Y_\Sigma^R$ are the conjugate
momenta for the twistor variables, and $F^\Sigma_R$ are auxiliary fields. The higher 
components are not independent fields and are required only for chirality.
 
The prepotential for the $U(4)$ symmetry has the following expansion
in Wess-Zumino gauge:
\begin{eqnarray}V^R_S = \sigma^R_S \kappa^+\bar\kappa^+ + \bar\sigma^S_R
\kappa^-\bar\kappa^+ + (A_\pp)^R_S\kappa^+\bar\kappa^+ + (A_{\mm})^R_S
\kappa^-\bar\kappa^- + \cdots ,
\end{eqnarray}
where the ellipsis contains the gauginos and higher 
components. Note that in this gauge 
\begin{eqnarray}
e^V= 1 + V + \half V^2,
\end{eqnarray}
where all terms above are matrices.\footnote{To avoid
cumbersome notation, we sometimes omit global $\Sigma$, local $R$, or both indices.} 
The prepotential has twisted-chiral field strengths given by
\begin{eqnarray}
\label{twistedchiral}
\Sigma \doteq \{ \bar\CD_+ ,\CD_-\}=\bar D_+ (e^{-V}D_-e^V)&,& \tilde\Sigma \doteq\{
\bar\CD_-,\CD_+\}= \bar D_- (e^{-V}D_+ e^V), 
\end{eqnarray} 
where, in the gauge-chiral representation, the covariant derivatives are given by
\begin{eqnarray} \label{chiralrep}
\CD_\pm = e^{-V} D_\pm e^V,\quad \bar\CD_\pm = \bar D_\pm.
\end{eqnarray}
The above field strengths are related to the usual chiral field strength defined
in four-dimensional, $N=1$ theories by
\begin{eqnarray}
\label{usualfs}
W_-= \bar\CD_- \Sigma,\quad W_+ =\bar\CD_+ \tilde\Sigma.
\end{eqnarray}
Another utility of the twisted-chiral field strengths is the addition of
a twisted-chiral superpotential to the model. For the present case, only
a linear superpotential will be added
\begin{eqnarray} \label{superpot} 
{\cal W}(\Sigma) = \frac{{\bf t}}{2}\Sigma,\quad \tilde{\cal
W}(\tilde\Sigma)= \frac{\bar {\bf t}}{2}
\tilde\Sigma,\end{eqnarray}
where ${\bf t}= t + i \frac{\theta}{2\pi}$, $t$ will represent the
squared radius of the sigma model, and $\theta$ couples to the first Chern
class of the gauge field. Unlike bosonic Grassmannian sigma models, there are
no dynamical corrections to this superpotential \cite{quantasp}. 

The action for this model is given by\footnote{This type of gauged linear
sigma model for Grassmannian manifolds was discussed in
\cite{linds}.}
\begin{eqnarray}
\label{action}
S&=&\int d^2\sigma d^4\kappa \Big[ \bar\Phi_\Sigma e^{V} \Phi^\Sigma
+ \frac{1}{g^2}\Tr( \Sigma \tilde\Sigma) \Big] 
	+ \int d^2\sigma d\kappa^+d\bar\kappa^-\frac{{\bf t}}{2}\Tr(\Sigma)  
+ \int d^2\sigma d\bar\kappa^+d\kappa^-\frac{\bar {\bf
t}}{2} \Tr(\tilde\Sigma), \nonumber \\
\end{eqnarray} where $g$ is the coupling constant for the gauge field with dimensions of $(\mathrm{length})^{-1}$.
Here, and in the rest of the paper, $\Tr(\cdot )$ 
denotes the trace over $U(4)$ indices.
The equations of motion for $\Phi^\Sigma_S$ and
$\bar\Phi^S_\Sigma$ with arbitrary $g$ and $\bf t$ are 
\begin{eqnarray}
\label{eqsmot}
D_+D_-[ (e^V)^R_S\Phi^\Sigma_R]=0,\quad \bar D_+\bar D_-
[\bar\Phi^R_\Sigma (e^V)^S_R]=0.
\end{eqnarray}

In the deep infra-red limit, $g\to \infty$, the equation of motion for $V$ that follows from this action is 
\begin{eqnarray}
\label{size}
t \delta^R_S= \bar\Phi^T_\Sigma(e^V)^R_T\Phi^\Sigma_S,\end{eqnarray}
whence we find that $t$ has an interpretation as the ``size'' of the super-Grassmannian
manifold. Another way to see this is to write the action as 
\begin{eqnarray}
S=\int d^2\sigma d^4\kappa [ \bar\Phi^R_\Sigma \Phi^\Sigma_R +
\frac{1}{2}\bar\Phi^R_\Sigma (V^2)_R^S \Phi^\Sigma_S + V^R_S(
\bar\Phi^S_\Sigma \Phi_R^\Sigma - t \delta^S_R) + \cdots],
\end{eqnarray} where the ellipsis denotes terms which vanish in Wess-Zumino gauge, and we
set $\theta=0$ and $g \to \infty$. We can
clearly see how the familiar constraint
$\bar\Phi^S_\Sigma\Phi^\Sigma_R=R^2\delta^S_R$ appears with $t=R^2$:
Besides being responsible for the gauge invariance, $V$ also plays the
role of the Lagrange multiplier in the $g\to \infty$ limit. It constrains the dynamical system defined by the action (\ref{action}) to the Grassmannian and is of a different nature than the differential equation of motion (\ref{eqsmot}). We will therefore distinguish the consequences of these two conditions by referring to equations holding due to (\ref{size}) as {\em off-shell} and those holding due to (\ref{eqsmot}) as {\em on-shell}.

The solution of equation (\ref{size}) is 
\begin{eqnarray} V^R_S = \delta^R_S \log t - \log(\bar\Phi^R_\Sigma
\Phi^\Sigma_S). \end{eqnarray}
Substituting this equation back into the action, we get a non-linear action in terms
of $(\Phi,\bar\Phi)$. Subsequently, using the $U(4)$ gauge invariance to
fix\footnote{Although useful, this gauge fixing is not very convenient
when one wants to study the relation between the GLSM and the pure spinor
version \cite{berkvafa,quantasp}.} 
$ \Phi^J_R=\sqrt{t}\delta^J_R $, we obtain
\begin{eqnarray}
\label{non-linear} S = t \int d^2z d^4\kappa \, \Tr \, [ \log( \delta^J_K +
\frac{1}{t}\bar\Phi^J_A\Phi^A_K)]
\end{eqnarray}
which is the usual $N=(2,2)$ non-linear sigma model action for Grassmannian
manifolds. 

We close this section with some comments on the interpretation of this 
gauged linear sigma model. The worldsheet supersymmetry is A-twisted,
which means that the components $(X^\Sigma_R,\bar X^R_\Sigma)$ of the (anti)chiral fields defined 
in equation (\ref{fundfields}) have conformal weight zero and the 
components $(\bar Y^\Sigma_R,Y^R_\Sigma)$ have conformal weight
one.\footnote{One should be careful when talking about conformal symmetry in
the present case since, as usual in gauged linear sigma models, the action
is only supposed to be conformaly invariant in the infrared limit.} However,
the worldsheet operators generating the superconformal transformations
are {\em not} the operators whose cohomology defines the physical spectrum.
This fact is due to the nontrivial mapping \cite{berkvafa} between the
pure spinor variables and the variables in equation (\ref{fundfields}). This mapping,
which involves two tensors $(\epsilon_{AB},\epsilon_{JK})$ (in addition to those defined in appendix \ref{Sec:Supergroups}) which explicitly break the
$U(2,2|4)$ symmetry, breaks worldsheet supersymmetry. In
conclusion, although the action (\ref{non-linear}) is topological in the
sense that it can be written in a BRST-exact form, the
spectrum and correlation functions are not those of a topological theory.

%%%%%%%%%%%%%%%%%%%%%%%%%%%%%%%%%%%%%%%%%%%%%%%%%%
\section{Classical Symmetries}
\label{symmetries}

In this section we analyze the symmetries of the action (\ref{action}). 
Our goal is to verify that this two-dimensional field theory 
is integrable at both the classical and quantum level. Although the
interpretation of the model is subtle, since it involves a field
redefinition of the standard worldsheet variables, we have a well-defined field theory in two dimensions, and it is worthwhile to study its properties. Little is known about sigma models on supergroup manifolds. It was shown in
\cite{grassmann} that the pure bosonic Grassmannian sigma model is not
integrable at the quantum level but its $N=1$ supersymmetric extension is. We
would like to know the analogous statement for the present model.

When $t\neq 0$ and $g\to \infty$ we can
integrate $V$ out and get a non-linear sigma model
(\ref{non-linear}) \cite{berk}. When $t=0$
this procedure cannot be carried out. It would be interesting to analyze both
cases, but since the latter does not appear to have a clear geometric
interpretation, we will restrict our attention to the case $t\neq 0$ in this work.

Let us first analyze the local and global symmetries of equation (\ref{action}). The
$U(4)$ gauge transformations are given by
\begin{eqnarray}
\delta \Phi^\Sigma_R &=& \delta L^S_R \Phi^\Upsilon_S,\quad \delta
\bar\Phi^R_\Sigma = (\delta L^\dagger)_S^R \bar\Phi_\Upsilon^S=
-\delta L^R_S \bar\Phi_\Upsilon^S, \cr
%~\cr
\delta (e^V)_S^R &=& \delta L^R_T (e^V)^T_S - (e^V)^R_T \delta L^T_S, \cr
%~\cr
\delta \Sigma^R_S &=& \delta L^R_T \Sigma^T_S - \Sigma^S_T \delta L^T_S,
\end{eqnarray}
where $\delta L^R_S$ is the parameter for the $U(4)$ gauge transformation. We can see
more clearly the invariance of the action using matrix notation:
\begin{eqnarray}
\delta \Phi^\Sigma &=& \delta L\Phi^\Sigma,\quad \delta \bar\Phi_\Sigma= -
\bar\Phi_\Sigma \delta L, \cr
\delta e^V&=& [\delta L,e^V],\cr
\delta \Sigma &=& [\delta L,\Sigma]. 
\end{eqnarray}
The action (\ref{action}) also has global $U(2,2|4)$ invariance
\begin{eqnarray}
\delta_{\rm global} \Phi_R &=& \delta M\Phi_R,\quad \delta_{\rm
global}\bar\Phi^R=-\bar\Phi^R \delta M,\cr
\delta_{\rm global} e^V&=&0,\quad \delta_{\rm global} \Sigma =0,
\end{eqnarray}
where $\delta M$ is the parameter for the global $U(2,2|4)$ transformation. To
compute the conserved current associated with this global symmetry, we promote
the parameter of the transformation for $\Phi$ to a chiral superfield
$\delta M$ and
the one for $\bar\Phi$ to an antichiral superfield $\delta \bar M$. The variation of the
action is 
\begin{eqnarray}
\delta S= \int d^2z d^4\kappa [ -\bar\Phi^S \delta \bar M (e^V)_S^R \Phi_R +
\bar\Phi^S (e^V)^R_S \delta M\Phi_R ],
\end{eqnarray}
this variation is zero when $\delta M=\delta \bar M$, that is, when $M$ is a constant
superfield. Varying with respect to to $\delta \bar M$ we get
\begin{eqnarray}
\frac{\delta S}{\delta \bar M^\Upsilon{}_\Sigma}=
%\int d^2zd^2\bar\kappa\, [
-(-1)^{|\Upsilon||\Sigma|}D_+D_-(\bar\Phi^S_\Upsilon (e^V)_S^R \Phi_R^\Sigma),
%],
\end{eqnarray}
so $ D_+D_-(\bar\Phi^S_\Upsilon (e^V)_S^R \Phi_R^\Sigma)=0$ is the
conservation law associated with the global invariance. As usual, the
conservation law is only valid on-shell (eq.~\ref{eqsmot}). We will define
the corresponding gauge invariant conserved current as 
\begin{eqnarray}
\label{current}
J^\Sigma{}_\Upsilon\doteq (-1)^{|\Sigma||\Upsilon|}\bar\Phi^S_\Upsilon (e^V)_S^R\Phi_R^\Sigma,
\end{eqnarray}
where $J$ is a hermitian matrix-valued (indeed, $\mathfrak u(2,2|4)$-valued) superfield which is linear:
\begin{eqnarray}
\label{linearJ}
D_+ D_- J^\Upsilon{}_\Lambda=0\quad \textrm{(on-shell)}.
\end{eqnarray}
Due to the $V$ equation of motion (\ref{size}), the super-trace of this $\mathfrak u(2,2|4)$ current gives the diameter (squared) 
\begin{eqnarray}
\label{Diameter}
(-1)^{|\Sigma|}J^\Sigma{}_\Sigma = 4t
\end{eqnarray}
of the Grassmannian manifold. 
Finally, the conserved charge is
\begin{eqnarray}
\label{consecharge}
Q^\Sigma{}_\Upsilon = \int d\sigma \left[\int d\kappa^+d\bar\kappa^+
J^\Sigma{}_\Upsilon +
\int d\kappa^-d\bar\kappa^- J^\Sigma{}_\Upsilon\right].
\end{eqnarray}
The vector components, given by
\begin{eqnarray}\label{vectorJ}
(J_\pp)^\Sigma{}_\Upsilon \doteq  [D_+,\bar D_+] J^\Sigma{}_\Upsilon,\quad
(J_\mm)^\Sigma{}_\Upsilon \doteq[D_-,\bar D_-] J^\Sigma{}_\Upsilon .
\end{eqnarray}
can be used to write this charge simply as $Q^\Sigma{}_\Upsilon= \int d\sigma J_\tau^\Sigma{}_\Upsilon. $ In this formula, and all such formul\ae{} for charges appearing henceforth, we take only the lowest component of each superfield on the right-hand-side of the equation.

Since the worldsheet spinors prefer lightcone coordinates, it is convenient for the execution of superspace manipulations to work in this basis. The lightcone time will be taken to be $\sigma^\mm=\frac12(\tau-\sigma)$. Then, the lightcone charge is given by $Q_\mathrm{lc}=\int \mathrm d \sigma^\pp J_\pp$ and conservation $\partial_\mm Q_\mathrm{lc}=0$ 
follows from the identity 
%$ -2i(-\partial_\tau \triangle_\tau +\partial_\sigma \triangle_\sigma) = [D_+D_-, \bar D_+ \bar D_-]$ 
$ i\partial_\pp [D_-,\bar D_-]+i\partial_\mm [D_+,\bar D_+] = [D_+D_-, \bar D_+ \bar D_-]$ 
and linearity (\ref{linearJ}) of $J$. 

%%%%%%%%%%%%%%%%%%%%%%%%%%%%%%%%%%%%%%%%%%%%%%%%%
\section{Classical Integrability}
\label{classint}
Besides the global symmetry described above, the action (\ref{action}) admits
non-local symmetries. This ought to be true, at least classically, since
the gauged linear sigma model is related by a field redefinition to the
pure spinor string in $AdS_5\times S^5$, and the latter has
non-local charges \cite{flat}. Validity of this description of the pure spinor string in quantum theory requires that these symmetries are not anomalous \cite{berk1}.
%Also, if this description is consistent with the pure spinor description, it should be true quantum mechanically \cite{berk1}. 
The existence of an infinite number of conserved charges is regarded as an indication that the model is
integrable. 
In this section, we will show how the first non-local charge is constructed from the $\mathfrak u(2,2|4)$ current $J$. We then construct the superspace Lax operators generating the complete set of non-local charges and explain the connection to the more familiar component analysis \cite{grassmann}.

%%%%%%%%%%%%%%%%%%%%
\subsection{Classical Non-local Charge}
\label{ClassicalCharge}
An interesting property of the current (\ref{current}) is the identity (valid off-shell when $g\to
\infty$) 
\begin{eqnarray}
\label{fund}
J^\Sigma{}_\Upsilon J^\Upsilon{}_\Theta = -t J^\Sigma{}_\Theta,
\end{eqnarray}
which holds due to equation (\ref{size}) and the definition (\ref{current}). For ease of reference, we will call this equation the ``first fundamental $J$-identity''.
Although this equation looks like
an ordinary algebraic equation, we have to remember that the superfields
$\Phi$ and $\bar\Phi$ are constrained ({\it viz.}~chiral). 
%As it will be argued below, this
%identity, together with two other identities to be discussed, is the analog of the flatness condition in two-dimensional
%classical integrable models. Using this equation, we will construct a
%superspace non-local charge.

%Before proving that, we derive some identities which will be useful in
%constructing the first non-local charge. 
We now derive the two remaining identities.
Multiplying equation (\ref{size})
on left by $\Phi^\Upsilon_R$ we obtain an off-shell identity which, together with its conjugate, can be written as
\begin{eqnarray}
J^\Upsilon{}_\Sigma \Phi^\Sigma_S=-t\Phi^\Upsilon_S  &\mathrm{and}& \bar \Phi^S_\Sigma J^\Sigma{}_\Upsilon=-t \bar \Phi^S_\Upsilon.
\end{eqnarray}
Applying $\bar D_\pm$ on the first equation and using chirality, one obtains $(\bar D_\pm J)\Phi =0$. Taking the complex conjugate of this equation gives $\bar\Phi (D_\pm J)=0$. These two equations imply the second and third fundamental identities\footnote{Although we give them different names for easy of reference, the third identity is the hermitian conjugate of the second.}
\begin{eqnarray}
\label{basi}
 (\bar D_\pm J^\Upsilon{}_\Sigma)J^\Sigma{}_\Lambda=0
	&\mathrm{and}&
(-1)^{|\Sigma|}J^\Upsilon{}_\Sigma (D_\pm J^\Sigma{}_\Lambda ) =0.
\end{eqnarray}
These two equations together with (\ref{fund}) will form the basic set of
fundamental off-shell equations. They represent the superspace analogue of the  
flatness condition in two-dimensional classical integrable models.
Combined with the on-shell relation
(\ref{linearJ}), they are, in fact, equivalent to the $V$ equation of
motion, and chirality and equations of motion of $\Phi$, thus providing the necessary ingredients to construct the flat component current and Lax operators.

To write the component equation for the curl of the conserved current, it is useful to define a second component current constructed from fermion bi-linears:
\begin{eqnarray}
\label{jbilinear}
j_{\pp}^\Upsilon{}_\Lambda&=& -\frac2t (-1)^{|\Upsilon|+|\Sigma|}\left( 
	D_+ J^\Upsilon{}_\Sigma \bar D_+ J^\Sigma{}_\Lambda 
	+\bar D_+ J^\Upsilon{}_\Sigma  D_+ J^\Sigma{}_\Lambda \right) ,\cr
j_{\mm}^\Upsilon{}_\Lambda&=& -\frac2t (-1)^{|\Upsilon|+|\Sigma|}\left( 
	D_- J^\Upsilon{}_\Sigma \bar D_- J^\Sigma{}_\Lambda 
	+\bar D_- J^\Upsilon{}_\Sigma  D_- J^\Sigma{}_\Lambda \right).
\end{eqnarray}
In appendix \ref{flatness} we show that these currents, together with $J_{\pp,\mm}$ satisfy the ``flatness equation'' \cite{grassmann}
\begin{eqnarray}
\label{flat}
it\partial_{\pp} \left( J_{\mm}^\Upsilon{}_\Lambda + j_{\mm}^\Upsilon{}_\Lambda \right) 
	- it\partial_{\mm}\left(  J_{\pp}^\Upsilon{}_\Lambda +j_{\pp}^\Upsilon{}_\Lambda \right)
	 + \left[ J_{\pp}^\Upsilon{}_\Sigma J_{\mm}^\Sigma{}_\Lambda - J_{\mm}^\Upsilon{}_\Sigma J_{\pp}^\Sigma{}_\Lambda\right] 
	 =0~.
\end{eqnarray}
With this, we are able to write down the non-local
charge. 
In lightcone coordinates,
\begin{equation}
\label{non-localcharge}
{\cal Q}_\mathrm{lc}= \int\!\!\!\!\int {\rm d}\sigma^\pp_1{\rm d}\sigma^\pp_2
\theta(\sigma^\pp_1-\sigma^\pp_2)[ J_\pp(\sigma_1),J_\pp(\sigma_2)]
	+2it\int{\rm d}\sigma^\pp (J_\pp + j_\pp).
\end{equation}
Using (\ref{flat}) and the conservation of the vector components of the
currents, it is straightforward  to verify that 
\begin{eqnarray}
\partial_\mm {\cal Q}_\mathrm{lc}=0,
\end{eqnarray}
that is, the non-local charge is conserved.

%%%%%%%%%%%%%%%%%%%%%%
\subsection{The Superspace Lax ``quartet''}
\label{Lax}

In this section we construct a superspace Lax representation of the
flatness equation. The starting point is to construct the following 
``pure gauge connections''
\begin{eqnarray}\label{superlax}
\mathbf D_+^\Upsilon{}_\Lambda =(-1)^{|\Sigma|} ( e^{-\frac\lambda t J})^\Upsilon{}_\Sigma D_+ \, (e^{\frac\lambda t J})^\Sigma{}_\Lambda
	&,& 
		\bar{\mathbf D}_+^\Upsilon{}_\Lambda= (-1)^{|\Sigma|} (e^{\frac\lambda t J} )^\Upsilon{}_\Sigma \bar D_+ \,( e^{-\frac\lambda t
J})^\Sigma{}_\Lambda,\cr
\mathbf D_-^\Upsilon{}_\Lambda =(-1)^{|\Sigma|} ( e^{\frac\lambda t J})^\Upsilon{}_\Sigma D_- \, (e^{-\frac\lambda t J})^\Sigma{}_\Lambda
	&,& 
		\bar{\mathbf D}_-^\Upsilon{}_\Lambda= (-1)^{|\Sigma|} (e^{-\frac\lambda t J} )^\Upsilon{}_\Sigma \bar D_- \,( e^{+\frac\lambda t
J})^\Sigma{}_\Lambda.
\end{eqnarray}
From these definitions it is clear that $\{\mathbf D_+,\bar\mathbf
D_-\}=\{\mathbf D_-,\bar\mathbf D_+\}=0$. To check which other
supercurvatures are zero we 
expand the derivations
(\ref{superlax}). 
Due to the first fundamental equation (\ref{fund}), it is easy to compute that
\begin{equation}
(e^{\frac{\lambda}{t}J})^\Sigma{}_\Lambda= \delta^\Sigma{}_\Lambda +
\frac{1}{t}(1-e^{-\lambda})J^\Sigma{}_\Lambda
\end{equation}
and, therefore, the derivations are at most quadratic in $J$. We now show
that they are, in fact, linear.
Explicitly we have
\begin{eqnarray}
\mathbf D_+^\Sigma{}_\Theta
&=&(-1)^{|\Sigma|}\delta^\Sigma{}_\Theta
D_+ + \frac{1}{t}(-1)^{|\Lambda|}(1-e^{-\lambda})(\delta^\Sigma{}_\Lambda +
\frac{1}{t}(1-e^{\lambda})J^\Sigma{}_\Lambda) D_+ J^\Lambda{}_\Theta.  
\end{eqnarray} 
We can simplify this by using the third fundamental relation (\ref{basi}) to obtain
\begin{equation}
\mathbf D_+^\Sigma{}_\Theta = (-1)^{|\Sigma|}\left( \delta^\Sigma{}_\Theta D_+ +
\frac{1}{t}(1-e^{-\lambda}) D_+ J^\Sigma{}_\Theta \right).
\end{equation}
The construction of $\bar\mathbf D_+$ involves an additional step:
At first we have
\begin{eqnarray}
\bar\mathbf D_+^\Sigma{}_\Theta &=& (-1)^{|\Sigma|}\delta^\Sigma{}_\Theta \bar D_+ +\frac{1}{t}(-1)^{|\Lambda|}(1-e^{\lambda})(\delta^\Sigma{}_\Lambda +
\frac{1}{t}(1-e^{-\lambda})J^\Sigma{}_\Lambda) \bar D_+ J^\Lambda{}_\Theta
\\
&=& (-1)^{|\Sigma|}\delta^\Sigma{}_\Theta \bar D_+ +
\frac{1}{t}(-1)^{|\Sigma|}(1-e^{\lambda})\bar D_+ J^\Sigma{}_\Theta +
\frac{1}{t^2}(2-e^\lambda -e^{-\lambda})\left(
(-1)^{|\Lambda|}J^\Sigma{}_\Lambda \bar D_+
J^\Lambda{}_\Theta\right).\nonumber
\end{eqnarray}
Now we simplify this using the second fundamental relation 
%$(\bar D_\pm J^\Sigma{}_\Lambda)J^\Lambda{}_\Theta=0$ 
(\ref{basi}) and the equation $(-1)^{|\Lambda|}J^\Sigma{}_\Lambda$ $\bar D_\pm J^\Lambda{}_\Theta= -t(-1)^{|\Sigma|} \bar D_\pm J^\Sigma{}_\Theta$, which follows from the first fundamental relation and is derived (c.f.~equation \ref{DJ}) in appendix \ref{flatness}.
This gives
\begin{equation}
\bar\mathbf D_+^\Sigma{}_\Theta=(-1)^{|\Sigma|}\left( \delta^\Sigma{}_\Theta \bar D_+
+\frac{1}{t}(e^{-\lambda}-1)\bar D_+ J^\Sigma{}_\Theta \right). 
\end{equation}
Along exactly the same lines, the remaining two derivations are giving by
\begin{eqnarray}
\mathbf D_-^\Sigma{}_\Theta&=& (-1)^{|\Sigma|}\left( \delta^\Sigma{}_\Theta D_-
+\frac{1}{t}(1-e^\lambda) D_- J^\Sigma{}_\Theta\right),\cr \cr
\bar\mathbf D_-^\Sigma{}_\Theta&=& (-1)^{|\Sigma|}\left( \delta^\Sigma{}_\Theta D_-
+\frac{1}{t}(e^\lambda-1)\bar D_- J^\Sigma{}_\Theta\right).
\end{eqnarray}

It is now easy to show that the supercurvature $\{ \mathbf D_+,\mathbf D_-\}$ vanishes 
if and only if all on- and off-shell $J$-identities hold. The same is true for
$\{ \bar\mathbf D_+,\bar\mathbf D_-\}$. 
%vanish due to the
%conservation law $D_+D_- J=0$ 
%and the second and third fundamental relations (\ref{basi}).
%The remaining two supercurvatures $\{\mathbf D_+,\bar\mathbf D_+\}$ and 
%$\{ \mathbf D_-,\bar\mathbf D_-\}$ cannot vanish, of course. We will
%return to this point shortly.
With these four derivations we define the compatible system of equations
\begin{equation}\label{newlaxeq}
{\mathbf D}_\pm U(\sigma^\pp,\sigma^\mm\, ;\lambda)=0\, ~~\mathrm{or}~~
\bar{\mathbf D}_\pm V(\sigma^\pp,\sigma^\mm\, ;\lambda)=0
\end{equation}
whose solutions generate infinitely many conservation laws. In order
to make contact with the usual Lax pair construction in bosonic integrable 
models we have to compute the two remaining supercurvatures. We will
define them as
\begin{equation}
{\mathbf D}_\pp = \frac{i}{2} \{ {\mathbf D}_+ ,\bar{\mathbf D}_+ \}~~\mathrm{and}~~
{\mathbf D}_\mm = \frac{i}{2} \{ {\mathbf D}_- ,\bar{\mathbf D}_- \}.
\end{equation}
Their explicit expressions can be computed using the equations above and the
result is
\begin{eqnarray}\label{oldlax}
\mathbf D_\pp^\Sigma{}_\Theta &=& \delta^\Sigma{}_\Theta\, \partial_\pp
+\frac{1}{2it}(1-e^{-\lambda})\,J_\pp^\Sigma{}_\Theta
-\frac{1}{4it}(1-2e^{-\lambda}+e^{-2\lambda})\, j_\pp^\Sigma{}_\Theta,
\cr\cr
\mathbf D_\mm^\Sigma{}_\Theta &=& \delta^\Sigma{}_\Theta\, \partial_\mm
+\frac{1}{2it}(1-e^\lambda)\, J_\mm^\Sigma{}_\Theta
-\frac{1}{4it}(1-2e^\lambda+e^{2\lambda})\, j_\mm^\Sigma{}_\Theta,
\end{eqnarray} 
where $(J_\pp,J_\mm)$ and $(j_\pp, j_\mm)$ were defined in equations (\ref{vectorJ}) and (\ref{jbilinear}), respectively.
Using the vanishing supercurvatures $\{ \mathbf D_+,\mathbf D_-\}$ and $\{ \bar\mathbf D_+,\bar\mathbf D_-\}$, we automatically have that 
\begin{equation}
\label{F0}
\mathbf F_{\pp\mm}(\lambda)\doteq [ \mathbf D_\pp, \mathbf D_\mm]=0
\end{equation}
which is the equation satisfied by the usual bosonic Lax pair. 

Expanding this expression in exponentials of the spectral parameter, we find
linearly independent combinations of $J$-flatness (\ref{flat}) and $J$-conservation, and analogous equations expressing the non-flatness and non-convervation of $j$. 
These formul\ae{} are equivalent to those found by component analysis in reference \cite{grassmann}
and the derivations in (\ref{oldlax}) correspond precisely to the usual Lax
pair in sigma models on Grassmannian manifolds.
It is known that solutions of 
\begin{equation}\label{oldlaxeq}
\mathbf D_\pp U(\sigma^\pp,\sigma^\mm\, ;\lambda) = \mathbf D_\mm
U(\sigma^\pp,\sigma^\mm\, ;\lambda)=0
\end{equation}
lead to infinitely many conservation laws \cite{Luscher:1977rq}. Of course every
solution of (\ref{newlaxeq}) with $V=U$ is also a solution of (\ref{oldlaxeq}).
Whether the reverse is true we leave as an interesting open question. 

%%%%%%%%%%%%%%%%%%%%%%%%%%%%%%%%%%%%%%%%%%%%%%%%%%
\section{Quantum Integrability}\label{quantumint}
The computations in section \ref{ClassicalCharge} relevant to the definition of the non-local charge (\ref{non-localcharge}) are a mixture of superspace and component calculations. To study the quantum analogue of the conservation of the non-local charge, one can proceed with the component analysis along the lines of reference \cite{Luscher:1977uq}. However, the non-local term in the charge is most easily proven to be unrenormalized by embedding it in superspace. We therefore prefer to keep supersymmetry manifest. Furthermore, since the worldsheet fermions $\kappa$ prefer lightcone coordinates, we will perform all calculations in this section in the lightcone basis.

%%%%%%%%%%%%%%%%%%%%%%
\subsection{Embedding of the Non-local Charge in Superspace}

We begin by proposing an $N=(2,2)$ generalization of the Heaviside function. This will be the formal substitution of the worldsheet supercoordinate in the ordinary Heaviside function. To construct the appropriate worldsheet supercoordinate we start with the chiral representation superspace lightcone coordinates $\sigma_1^\pp -\sigma_2^\pp +i \bar \kappa_2^+ \kappa_1^+$ and $\sigma_1^\mm -\sigma_2^\mm + i \bar \kappa_2^- \kappa_1^-$. These coordinates have the property that they are annihilated by $\bar D_{\pm1}$ and $D_{\pm2}$ in the chiral representation.
Since we will be working with hermitian superfields, it is appropriate to switch to the real representation\footnote{We thank Martin Ro\v{c}ek and Warren Siegel for reminding us of this expression.} obtained by acting with $\mathrm e^{i(\kappa\sigma^a\bar \kappa )\partial_a}$. This gives 
\begin{eqnarray}
\label{wscoords}
\hat \sigma_{12}^\pp&=& \sigma_1^\pp -\sigma_2^\pp +i \bar\kappa_2^+ \kappa_1^+ + i (\kappa_1^+\bar \kappa_1^+ - \kappa_2^+\bar \kappa_2^+), \cr
\hat \sigma_{12}^\mm&=&\sigma_1^\mm -\sigma_2^\mm + i \bar\kappa_2^- \kappa_1^- + i (\kappa_1^-\bar \kappa_1^- - \kappa_2^-\bar \kappa_2^-) .
\end{eqnarray}
With this expression for the worldsheet coordinate, the proposal for the Heaviside function is simply
\begin{eqnarray}
\Theta( \sigma^\pp_{12})\doteq \theta(\hat \sigma^\pp_{12}).
\end{eqnarray}
The na\"ive guess for the first term in the supercharge is the Lorentz covariant integral
\begin{eqnarray}
I_0= \int\!\!\!\! \int \mathrm d\mu_{1} \mathrm d \mu_{2} \, \Theta(\sigma^\pp_{12}) \left[ J( \sigma_1), J(\sigma_2)\right]
\end{eqnarray}
with the ``measure'' $ \mathrm d\mu = \mathrm d \sigma [D_+,\bar D_+] |$. To check this we must compute the component projection. To do that it is useful to notice that the Heaviside function depends only on even powers of $\kappa$. This, together with anti-symmetry of the commutator, results in only two non-vanishing contributions: One in which both commutators hit the Heaviside function and one in which neither of them do. Direct calculation results in
\begin{eqnarray}
\label{oops}
I_0= \int\!\!\!\! \int \mathrm d\sigma^\pp_1 \mathrm d \sigma^\pp_2 \,\Big\{\theta(\sigma^\pp_{12})\left[ J_\pp(\sigma_1), J_\pp(\sigma_2)\right]
	+ 4\delta^\prime(\sigma^\pp_{12}) \left[ J(\sigma_1), J(\sigma_2)\right]	\Big\}~.
\end{eqnarray}
Integrating the $\delta^\prime$ term over $\sigma_{1,2}$ we get
\begin{eqnarray}
-4 \int \!\!\!\! \int \mathrm d\sigma^\pp_1 \mathrm d \sigma^\pp_2 \,
	\delta(\sigma^\pp_1 -\sigma^\pp_2) \left[ \partial_\pp J (\sigma_1) , J(\sigma_2) \right] 
	= 4 \int \mathrm d \sigma J \stackrel{\leftrightarrow}{\partial_\pp} J.
\end{eqnarray}
Although this type of term does not look familiar, we show in appendix \ref{flatness} that
\begin{eqnarray}
\label{j}
J \stackrel{\leftrightarrow}{\partial_a} J=\frac{it}2(j_a-J_a),
\end{eqnarray}
where $a$ can be any of the indices $\pp, \mm, \tau, \sigma$. 
The superspace integral is therefore expressible as
\begin{eqnarray}
I_0 =\int\!\!\!\! \int \mathrm d\sigma^\pp_1 \mathrm d \sigma^\pp_2 \,\theta(\sigma^\pp_{12})\left[ J_\pp(\sigma_1), J_\pp(\sigma_2)\right]
	+ 2it \int \mathrm d \sigma^\pp \left( - J_\pp+ j_\pp \right).
\end{eqnarray}
It follows that the non-local charge in lightcone coordinates is expressible entirely in terms of a the $\mathfrak u(2,2|4)$ supercurrent as 
\begin{eqnarray}
\label{superQ}
\mathcal Q_\mathrm{lc} = \int\!\!\!\! \int \mathrm d\mu_{1} \mathrm d \mu_{2} \, \Theta(\sigma^\pp_{12}) \left[ J( \sigma_1), J(\sigma_2)\right] +4it \int \mathrm d\mu \, J.
\end{eqnarray}

The precise relative coefficient in the component expression (\ref{oops}) is crucial to match the coefficient of $j_\pp$ in (\ref{non-localcharge}). This is important since, contrary to $J_\pp = [D_+,\bar D_+] J$, it is impossible to write $j_\pp$ as an expression of the form $\left((\mathrm{combination~of~}D_+~\mathrm{and}~\bar D_+)\mathrm{~acting~on~}(\mathrm{function~of~}J)\right)$. It would have followed from this that there is no superspace expression, the lowest component of which is the non-local charge. 
%Alternatively, the superspace form (\ref{superQ}) fixes the relative coefficient of the two terms in the component form of the non-local charge (\ref{non-localcharge}).

%%%%%%%%%%%%%%%%%%%%%%%%%%%
\subsection{Non-renormalization}
We now examine the possible renormalization of the the non-local $\mathfrak u(2,2|4)$ charge. 
%Since much of what we will say is true for any Lie superalgebra $\mathfrak g$, we will only take $\mathfrak g = \mathfrak u(2,2|4)$ when necessary.
The superspace form (\ref{superQ}) shows that if the charge is renormalized, it will happen 
due to the operator product of the $\mathfrak u(2,2|4)$ currents $J(\sigma_1)$ and $J(\sigma_2)$. We will now show that the supergroup nature of this operator product cancels this potential divergence.

Consider, again, the $\mathfrak u(2,2|4)$ current $J^\Upsilon{}_\Sigma$.
% for the Lie super-algebra $\mathfrak g$ of some general Lie supergroup $G$.
Irrespective of how it is defined, its operator product expansion with
any operator ${\cal O}^\Theta$ in the fundamental representation is on general grounds
\begin{eqnarray}\label{ope1}
J^\Upsilon{}_\Sigma(\hat \sigma_1){\cal O}^\Theta(\hat \sigma_2)\sim -\,
\log|\hat \sigma_{12}|^2 \,T_{\Sigma\Gamma}^{\Upsilon \Theta}\,{\cal
O}^\Gamma(\hat \sigma_{1+2})
\end{eqnarray}
for some $\mathfrak u(2,2|4)$-invariant tensor $T$. Where we assumed that ${\cal
O}^\Theta$ is a chiral operator. This OPE is
constrained by the fact that ${\cal O}^\Theta$ must transform under a global $U(2,2|4)$
transformation as 
\begin{equation}
\label{FundRep}
[ M^\Sigma{}_\Upsilon {Q}^\Upsilon{}_\Sigma \, ,{\cal O}^\Theta ]=
M^\Theta{}_\Upsilon {\cal O}^\Upsilon,
\end{equation}
where $Q^\Upsilon{}_\Sigma$ is given by (\ref{consecharge}).
This form is also fixed by the classical weight of $J$, which does not
change in the quantum theory since $J$ is a conserved current. 
The tensor structure is determined by the action of 
%$\mathfrak g$ which for the fundamental representation is given by
$\mathfrak u(2,2|4)$.
In appendix \ref{Appendix} we review the construction of the
$\mathfrak{u}(2,2|4)$ algebra.
There we find that 
$T_{\Sigma\Gamma}^{\Upsilon
\Theta}=(-1)^{|\Theta|} \delta^\Theta{}_\Sigma \delta^\Upsilon{}_\Gamma$
is the sign factor (\ref{fundrep}).
%as we review explicitly (c.f.~equation (\ref{fundrep})) for the case $\mathfrak g=\mathfrak u(2,2|4)$ in appendix \ref{Appendix}. 
%One can check that (\ref{ope1}) indeed satisfy these
%requirements by computing explicitly the OPE of the vector component of
%$J$ with $\cal O$. {\bf I don't understand this statement.}

The $JJ$ operator product now follows from the adjoint action and also by
acting $J$ twice in (\ref{ope1})
%of $\mathfrak g$ on itself. For $\mathfrak g=\mathfrak u(2,2|4)$ (c.f.~equation (\ref{JJ})),
\begin{eqnarray}
\label{AdjRep}
&&\hspace{-1.5cm}  J^\Upsilon{}_\Sigma(\hat \sigma_1) J^\Theta{}_\Phi(\hat \sigma_2)\sim \cr 
&& \hspace{-1cm} -(\,\log|\hat \sigma_{12}|^2 +\log|\hat\sigma_{21}|^2\,) 
	\Big[ (-1)^{|\Phi|(|\Sigma|+|\Theta|)+|\Sigma||\Theta|}
\delta^\Upsilon{}_\Phi J^\Theta{}_\Sigma(\hat \sigma_{1+2}) -
(-1)^{|\Sigma||\Theta|}\delta^\Theta{}_\Sigma J^\Upsilon{}_\Phi(\hat \sigma_{1+2})
	\Big]
\end{eqnarray} where we have included $\log|\hat\sigma_{21}|^2$ so that
this OPE is hermitian. Note that $\hat\sigma_{12}$ is {\it not}
antisymmetric in 1 and 2.\footnote{Terms like
$\frac{\sigma^\pp}{\sigma^\mm}$ are also forbidden in these OPEs since
$J$ is a worldsheet scalar.}
What enters the quantum charge, however, is the matrix product. This
corresponds to summing over $\Theta=\Sigma$. This gives (c.f.~equation (\ref{JJprod}))
\begin{eqnarray}
J^\Upsilon{}_\Sigma(\hat\sigma_1)J^\Sigma{}_\Phi(\hat\sigma_2) \sim
-4t(\log|\hat\sigma_{12}|^2
+\log|\hat\sigma_{21}|^2)\delta^\Upsilon{}_\Phi
%(-1)^{|\Sigma|}J^\Sigma{}_\Sigma
,
\end{eqnarray}
where we have used equation (\ref{Diameter}) for the super-trace of $J$.
%but as was shown in section \ref{symmetries}
%$(-1)^{|\Sigma|}J^\Sigma{}_\Sigma=4t$. 
Finally, what enters the quantum non-local charge is
% we are interested in
the commutator%, which vanishes
\begin{eqnarray}
J^\Upsilon{}_\Sigma(\hat \sigma_1)J^\Sigma{}_\Phi(\hat
\sigma_2)-J^\Upsilon{}_\Sigma(\hat\sigma_2)J^\Sigma{}_\Phi(\hat\sigma_1)\sim
0
\end{eqnarray}
which is, therefore, not renormalized. All other potential quantum
corrections to the equation above are of order $|\sigma|^2$ since the gauge coupling 
constant and other gauge invariant operators have negative length dimension. 

One can calculate from this OPE the corresponding
OPEs of the vector components of $J$, and they all vanish. 
This result has two consequences. First,
it means that the classical non-local charge (\ref{non-localcharge}) is well
defined in the quantum theory. Also, as a quantum operator, it is
conserved since all equations needed to prove this do not receive quantum corrections.

We have explicitly worked out the details of the operator products for the
$\mathfrak u(2,2|4)$ current of the $U(2,2|4)/U(2,2)\times U(4)$ and shown
that the non-local charge constructed from this current is not
renormalized. This result holds in more generality. Let us replace
$U(2,2|4)$ with a general supergroup $G$ with Lie super-algebra $\mathfrak
g$. Let $H\subset G$ be a subgroup and $K\subset G$ its commutant. The
gauged linear sigma model on the Grassmannian manifold $G/(K\times H)$ can
be constructed along the lines section \ref{Definition}, the $H$-invariant
current as in section \ref{symmetries}, the non-local
$G$-charge by the results of section \ref{classint}, and finally, its embedding in
superspace performed in this section. What is then required is to repeat
the steps considered here to show that the generator of this non-local
symmetry is not renormalized if the last term on the right-hand-side 
of (\ref{AdjRep}) vanishes. The operator product expansions entering
this calculation are, again, fixed by conformal weights and the
representation theory of $\mathfrak g$. Since we use only the fundamental
and adjoint representations, equations analogous to (\ref{FundRep}) and
(\ref{AdjRep}) hold. In the final step we take the matrix product of the
currents. The coefficient of the resulting operator product is simply the
generalization of the dual Coxeter number to the Lie super-algebra
$\mathfrak g$. We, therefore, conclude that any $N=(2,2)$ non-linear sigma
model with Grassmannian target manifold constructed from a supergroup $G$
with vanishing dual Coxeter number has a non-local $G$-symmetry which is
protected from renormalization.

In the case of no supersymmetry on the worldsheet, the Grassmannian sigma
model has an anomaly ({\it i.e.}~a gauge field strength appearing on the right-hand-side 
of (\ref{AdjRep}))  which prevents the non-local charge from being conserved
\cite{grassmann,abdalla2}. This anomaly disappears on the $N=(1,1)$
supersymmetric worldsheet which can also be seen from the fact that the
dimension of the
supersymmetric field strength prevents it from appearing in the OPE of the
supercurrents.

The renormalization of the non-local charge in the case of non-vanishing
dual Coxeter number is intimately related the the existence of a mass gap
in the theory. Since the Berkovits-Vafa GSLM does not have a mass gap
\cite{quantasp} it is natural to find that the non-local charge is not
renormalized.

%%%%%%%%%%%%%%%%%%%%%%%%%%%%%%%%%%%%%%%%%%%%%%%%%%%
\section{Conclusions and Further Directions}
\label{Conclusion}

In this paper we analyzed the classical and quantum integrability properties of the gauged linear sigma model proposed for the pure spinor superstring in $AdS_5\times S^5$ background by Berkovits and Vafa \cite{berkvafa}. A superspace
non-local charge was constructed and was proven to be conserved at both
the classical and quantum level. Furthermore, we constructed a superspace
Lax ``quartet'', which could be used to study the integrability of the
model directly in superspace. However, the use of such nontrivial
conservation laws in the present model still remains to be uncovered.

There are many interesting directions which deserve further study. 
One outstanding problem is to understand the precise
mapping between physical deformations of the original pure spinor action
and physical deformations of the action (\ref{action}). As we noted in
the introduction and in section \ref{Definition} the mapping will break
worldsheet supersymmetry since the cohomology is not defined as that of
the usual A-models. It would be interesting to see whether the non-local charge
constructed here commutes with the pure spinor BRST charge, as in
\cite{berk4}.

Another important open problem is a careful analysis of
the $t\to 0$ limit. We can see from the results above that the present approach
fails in this limit, since the construction does not
work in this case (many expressions are singular for $t=0$). Moreover, in this limit, one cannot eliminate the gauge
degrees of freedom. It is reasonable to expect that a suitable combination
of limits of both coupling constants leads to the existence of some other
nontrivial conservation laws. 

In \cite{read} the spectra of some coset sigma models with target space
supersymmetry were computed. These sigma models can be thought of as
supersymmetric generalizations of the $\vec n$ field model with a suitable 
topological term turned on, which in the present case, means a non-zero
$\theta$-angle in the superpotential (\ref{superpot}). One might wonder whether similar methods could be generalized for symmetric space cosets. 

%%%%%%%%%%%%%%%%%%%%%%%%%%%%%%%%%%%%%%%%%%%%%%%%%%%
\section*{Acknowledgements}

We would like to thank Elcio Abdalla, Nathan Berkovits, Vladimir Kazakov, Matthias Staudacher and Robert Wimmer
for useful discussions. We also thank Nathan Berkovits and Matthias
Staudacher for reading an earlier version of the draft and making useful
suggestions. The work of WDL3 is supported by NSF grant n$^{\rm o}$ PHY 0653342
and n$^{\rm o}$ DMS 0502267. 

%%%%%%%%%%%%%%%%%%%%%%%%%%%%%%%%%%%%%%%%%%%%%%%%%%%
%%%%%%%%%%%%%%%%%%%%%%%%%%%%%%%%%%%%%%%%%%%%%%%%%%%
\appendix

%%%%%%%%%%%%%%%%%%%%%%%%%%%%%%%%%%%%%%%%%%%%%%%%%%%
\section{The $U(2,2|4)$ Supergroup}
\label{Sec:Supergroups}
The supergroup $U(2,2|4)$ can be thought
as the group of unitary transformation of an eight-dimensional vector
space where the first four components are usual complex numbers and the
remaining four are complex Grassmann numbers. For example, let us denote
$X^\Sigma$ and element of this vector space. The index $\Sigma$ splits
into $A=1\cdots4$ and $J=1\cdots4$, {\it i.e.} $X^\Sigma=(x^A, \theta^J)$.
Here, $x^A$ are complex numbers and $\theta^J$ are complex Grassmann
numbers. The metric in this space is
$\eta_{\bar\Sigma\Upsilon}=(\eta_{\bar A B},\eta_{\bar I J})$, with
$\eta_{\bar A J}=\eta_{\bar I B}=0$. Furthermore, $\eta_{\bar A B}={\rm diag}(1,1,-1,-1)$
and $\eta_{\bar I J}={\rm diag}(1,1,1,1)$. The elements of $U(2,2|4)$ preserve
the inner product
\begin{eqnarray} \bar Y^{\bar\Sigma} \eta_{\bar\Sigma \Upsilon }X^\Upsilon=\bar Y_\Sigma
X^\Sigma=
(\bar Y')_\Sigma( X')^\Sigma,\end{eqnarray} where $(X')^\Sigma=
M^\Sigma{}_\Omega X^\Omega$, $(\bar Y')_\Sigma= \bar Y_\Omega
(M^\dagger)^\Omega{}_\Sigma$ and $M^\Sigma{}_\Omega$ is an element of
$U(2,2|4)$. Note that the supermatrix $M$ has the following form
\begin{eqnarray}
M^\Sigma{}_\Omega = \pmatrix{ m^A{}_B & f^A{}_J \cr g^I{}_B &
n^I{}_J },
\end{eqnarray}
where $m$ and $n$ are usual complex matrices and $f$ and $g$ are Grassmann
valued matrices. The conditions from invariance of the inner
product impose on these matrices are 
\begin{eqnarray}
(M^\dagger)^\Omega{}_\Sigma M^\Sigma{}_\Upsilon=\pmatrix{
(m^\dagger)^A{}_B m^B{}_C + (g^\dagger)^A{}_J g^J{}_C &
(m^\dagger)^A{}_B f^B{}_K +(g^\dagger)^A{}_J n^J{}_K \cr & \cr
(f^\dagger)^I{}_B m^B{}_C +(n^\dagger)^I{}_J g^J{}_C & 
(f^\dagger)^I{}_B f^B{}_K +(n^\dagger)^I{}_J n^J{}_K } =\pmatrix{
\eta^A{}_C & 0 \cr & \cr 0 & \eta^I{}_K }.\nonumber \\
\end{eqnarray}
These conditions can be solved factorizing $M$ into two matrices
\begin{equation}
M^\Sigma{}_\Omega = T^\Sigma{}_\Upsilon U^\Upsilon{}_\Omega,
\end{equation}
where the matrices $U$ and $T$ are given by
\begin{eqnarray}
U^\Sigma{}_\Omega &=& \pmatrix{ u^A{}_B & 0 \cr 0 & v^I{}_J}, \\ \nonumber \\
T^\Sigma{}_\Omega &=& \pmatrix{ \left(\frac{1}{ \sqrt{1+ Z Z^\dagger}}\right)^A{}_B &
Z^A{}_J \left(\frac{1}{\sqrt{ 1 + Z^\dagger Z}}\right)^J{}_K \cr  & \cr 
\left(\frac{1} { \sqrt{1+ Z^\dagger Z}}\right)^I{}_J (Z^\dagger)^J{}_B & 
\left(\frac{1}{\sqrt{ 1+ Z^\dagger Z}}\right)^I{}_K },
\end{eqnarray}\\
where $u$ and $v$ are two arbitrary $U(2,2)$ and $U(4)$ matrices
respectively and $Z$ is an arbitrary complex Grassmann valued matrix.
%%%%%%%%%%%%%%%%%%%%%%%%%%%%%%%%%%%%%%%%%%%%%%%%%%
\section{The $\mathfrak{u} (2,2|4)$ Algebra}
\label{Appendix}

Here we describe the superalgebra $\mathfrak{u}(2,2|4)\cong \mathfrak{gl}(4|4)$. We use mostly 
the definitions and conventions of \cite{gunaydin,beisert1}. In order to describe
this algebra, we introduce the set of oscillators ${\bf A}^\Sigma=({\bf
a}^\alpha,{\bf b}^\dagger_{\dot\alpha},{\bf c}^J)$, where we have split
the $A$ index into $(\alpha,\dot\alpha)$. Also, we define the hermitian
conjugate to be ${\bf A}^\dagger_\Sigma=({\bf a}^\dagger_\alpha,-{\bf
b}^{\dot\beta},{\bf c}^\dagger_J)$. The oscillators $({\bf a},{\bf b})$
are bosonic and the oscillators ${\bf c}$ are fermionic. They satisfy the
following (anti-)commutation relations:
\begin{eqnarray}
\label{commurel}
[{\bf a}^\alpha,{\bf a}^\dagger_\beta]=\delta^\alpha_\beta,\quad 
[\bb^{\dot\alpha},\bbd_{\dot\beta}]=\delta^{\dot\alpha}_{\dot\beta},\quad 
\{ \cb^J,\cbd_K \} =\delta^J_K, 
\end{eqnarray}
so we can define the graded commutators of $\Ab$ and $\Abd$ as 
\begin{eqnarray}\label{gradcommu}
[ \Ab^\Sigma,\Abd_\Upsilon] \doteq \Ab^\Sigma \Abd_\Upsilon
-(-1)^{|\Sigma||\Upsilon|} \Abd_\Upsilon \Ab^\Sigma =
\delta^\Sigma_\Upsilon,
\end{eqnarray}
where $|\Sigma|=0,1$ is the grading of the corresponding mode of the
oscillator. Using the above definitions, the generators of $\mathfrak{u}(2,2|4)$ are written as 
\begin{eqnarray}
\label{generators}
\mathfrak{J}^\Sigma{}_\Upsilon \doteq (-1)^{|\Sigma||\Upsilon|} \Abd_\Upsilon \Ab^\Sigma.
\end{eqnarray}
Note that the supervector $\Ab$ forms a fundamental representation of
$\mathfrak{u}(2,2|4)$ in the sense that
\begin{eqnarray}
\label{fundrep}
[\mathfrak{J}^\Sigma{}_\Upsilon,\Ab^\Theta]=-(-1)^{|\Theta|}\delta^\Theta_\Upsilon
\Ab^\Sigma.
\end{eqnarray}
The (anti-)commutation relations of the generators can be easily computed
using (\ref{gradcommu})
\begin{eqnarray}
\label{JJ}
[\mathfrak{J}^\Sigma{}_\Upsilon,\mathfrak{J}^\Theta{}_\Lambda] =
(-1)^{|\Lambda|(|\Theta|+|\Upsilon|)+|\Theta||\Upsilon|}\delta^\Sigma_\Lambda \mathfrak{J}^\Theta{}_\Upsilon
-(-1)^{|\Upsilon|}\delta^\Theta{}_\Upsilon
\mathfrak{J}^\Sigma{}_\Lambda.
\end{eqnarray}

It is interesting to compute the above commutator with the indices
$\Upsilon$ and $\Theta$ contracted 
\begin{eqnarray}\label{JJprod}
[\mathfrak{J}^\Sigma{}_\Theta,\mathfrak{J}^\Theta{}_\Lambda] =
(-1)^{|\Theta|}\delta^\Sigma_\Lambda \mathfrak{J}^\Theta{}_\Theta
-\mathfrak{J}^\Sigma{}_\Lambda
( (-1)^{|\Theta|} \delta^\Theta{}_\Theta)=
-2\delta^\Sigma_\Lambda\mathfrak{C},
\end{eqnarray}
where $\mathfrak{C}=-\half(-1)^{|\Theta|}\mathfrak{J}^\Theta{}_\Theta=-\half
\Ab^\dagger_\Theta \Ab^\Theta$ is the central charge operator. The other
possible trace of the generators is $\mathfrak{J}^\Theta{}_\Theta=
(-1)^{|\Theta|}\Ab^\dagger_\Theta \Ab^\Theta=2\mathfrak{C}+4\mathfrak{B}$ where 
\begin{equation}
\mathfrak{B}= \frac{1}{4} \mathfrak{J}^\Theta{}_\Theta
+\frac{1}{4}(-1)^{|\Theta|}\mathfrak{J}^\Theta{}_\Theta
\end{equation} is the hypercharge
\cite{beisert1}. These two traced generators can be removed from the
$\mathfrak{u}(2,2|4)$ algebra, and the end result is the
$\mathfrak{psu}(2,2|4)$ algebra.

%%%%%%%%%%%%%%%%%%%%%%%%%%%%%%%%%%%%%%%%%%%%%%%%%%
\section{Derivation of the ``Flatness Equation'' and ``$J$-relation''}
\label{flatness}
In this section we derive the flatness equation (\ref{flat}) used to construct the non-local conserved charge $\mathcal Q$ (\ref{non-localcharge}) and the relation (\ref{j}) between $J$, $J_a$, and $j_a$.
As this will involve some superspace gymnastics, we introduce the following notational aid for the commutator of two superspace derivatives: $\triangle_{\alpha \dot \alpha}\doteq [D_\alpha, \bar D_{\dot \alpha}]$. The dotted index refers to a label on a conjugated superspace derivative.

Hitting the second and third fundamental identities (\ref{basi}) with $D$ and $\bar D$ we find the relations
\begin{eqnarray}
-i \partial_{\alpha \dot \alpha}J^\Upsilon{}_\Sigma J^\Sigma{}_\Upsilon+\frac 12 \triangle_{\alpha \dot 	\alpha}J^\Upsilon{}_\Sigma J^\Sigma{}_\Lambda 
	-(-1)^{|\Upsilon|+|\Sigma|} \bar D_{\dot \alpha}J^\Upsilon{}_\Sigma D_\alpha J^\Sigma{}_\Lambda
	&=&0,\cr
-i J^\Upsilon{}_\Sigma \partial_{\alpha \dot \alpha}J^\Sigma{}_\Upsilon-\frac 12 J^\Upsilon{}_\Sigma \triangle_{\alpha \dot \alpha} J^\Sigma{}_\Lambda 
		+(-1)^{|\Upsilon|+|\Sigma|} \bar D_{\dot \alpha}J^\Upsilon{}_\Sigma D_\alpha J^\Sigma{}_\Lambda
	&=&0, 
\end{eqnarray}
where we have temporarily resorted to four-dimensional spinor notation to avoid a proliferation of formul\ae{}. Summing these equations and using the first fundamental identity (\ref{fund}) yields
\begin{eqnarray}
\label{covconst}
it \partial_a J -\frac 12 J \stackrel{\leftrightarrow}{\triangle}_aJ=0
\end{eqnarray}
where $a=\pp, \mm$ or $a=\tau, \sigma$. Taking the difference, we find
\begin{eqnarray}
\label{inter1}
i J^\Upsilon{}_\Sigma \stackrel{\leftrightarrow}{\partial}_{\alpha \dot \alpha} J^\Sigma{}_\Lambda
	-\frac t2 \triangle_{\alpha \dot \alpha}J^\Upsilon{}_\Lambda-2 (-1)^{|\Upsilon|+|\Sigma|} \bar D_{\dot \alpha}J^\Upsilon{}_\Sigma D_\alpha J^\Sigma{}_\Lambda.
\end{eqnarray}
Next, we rearrange the second and third fundamental identity to give
\begin{eqnarray}
\label{DJ}
-t \bar D_{\dot \alpha} J^\Upsilon{}_\Lambda-(-1)^{|\Upsilon|+|\Sigma|} J^\Upsilon{}_\Sigma \bar D_{\dot 	\alpha} J^\Sigma{}_\Lambda=0 &\mathrm{and}&
-t D_{\alpha} J^\Upsilon{}_\Lambda- J^\Upsilon{}_\Sigma D_{\alpha} J^\Sigma{}_\Lambda=0.
\end{eqnarray}
Hitting the first with $D$ and the second with $\bar D$ we get some unilluminating equations. The sum of these again gives (\ref{covconst}) but the difference gives
\begin{eqnarray}
i J^\Upsilon{}_\Sigma \stackrel{\leftrightarrow}{\partial}_{\alpha \dot \alpha} J^\Sigma{}_\Lambda
	-\frac t2 \triangle_{\alpha \dot \alpha}J^\Upsilon{}_\Lambda-2 (-1)^{|\Upsilon|+|\Sigma|} D_{\alpha}J^\Upsilon{}_\Sigma \bar D_{\dot \alpha} J^\Sigma{}_\Lambda.
\end{eqnarray}
Adding this to the intermediate result (\ref{inter1}) and taking the definition of the bi-linear current (\ref{jbilinear}) into account, we obtain the formula (\ref{j}) relating $J$, $J_a$, and $j_a$.

We now turn to the flatness equation (\ref{flat}). In this computation we take the formula (\ref{covconst}) and hit it with $\triangle_{\beta\dot \beta}$. We are interested in the case in which $a=\pp$ and $b=\mm$ or {\it vice versa}. The corresponding $D$s anti-commute and terms with 3 $D$s can be rewritten using the linearity of J. Taking all of this into account the formula simplifies to
\begin{eqnarray}
\triangle_b(it \partial_a J^\Upsilon{}_\Lambda)&=& \frac 12\left(  \triangle_b J^\Upsilon{}_\Sigma \triangle_aJ^\Sigma{}_\Lambda - \triangle_a J^\Upsilon{}_\Sigma \triangle_bJ^\Sigma{}_\Lambda\right)
	-it \partial_a j_b^\Upsilon{}_\Lambda\cr
	&+&\frac 12\left(J^\Upsilon{}_\Sigma \triangle_b \triangle_aJ^\Sigma{}_\Lambda - \triangle_b\triangle_a J^\Upsilon{}_\Sigma J^\Sigma{}_\Lambda\right).
\end{eqnarray}
Switching $a$ and $b$ and subtracting cancels the second line and gives the desired relation (\ref{flat}).

%{\footnotesize
%%%%%%%%%%%%%%%%%%%%%%%%%%%%%%%%%%%%%%%%%%%%%%%%

%} % This closes the footnotesize command above

\end{document}